\documentclass[aps,twocolumn,superscriptaddress,nofootinbib]{revtex4-2}

\pdfoutput=1

\usepackage{graphicx}
\usepackage{amsmath}
\usepackage{comment}
\usepackage[caption=false]{subfig}
\usepackage[colorlinks, citecolor={blue!70!black}, urlcolor={blue!70!black}, linkcolor={red!70!black},hyperindex,breaklinks]{hyperref}

\usepackage{xcolor}

\DeclareMathOperator{\Tr}{Tr}

\begin{document}

\title{A stability bound on the $\mathbf T$-linear resistivity of conventional metals}

\author{Chaitanya Murthy}
\affiliation{Department of Physics, Stanford University, Stanford, CA 93405, USA}

\author{Akshat Pandey}
\affiliation{Department of Physics, Stanford University, Stanford, CA 93405, USA}

\author{Ilya Esterlis}
\affiliation{Department of Physics, Harvard University, Cambridge MA 02138, USA}

\author{Steven A.~Kivelson}
\affiliation{Department of Physics, Stanford University, Stanford, CA 93405, USA}

\date{\today}

\begin{abstract}
Perturbative considerations account for the  properties of conventional metals, including the range of temperatures where the transport scattering rate is $1/\tau_\text{tr} = 2\pi \lambda T$, where $\lambda$ is a dimensionless  strength of the  electron-phonon coupling.
The fact that measured values satisfy $\lambda \lesssim 1$ has been noted in the context of a possible ``Planckian'' bound on transport. 
However, since the electron-phonon scattering is quasi-elastic in this regime, no such Planckian considerations can be relevant. 
We present and analyze Monte Carlo results on the Holstein model which show that a different sort of bound is at play: a ``stability'' bound on $\lambda$ consistent with metallic transport. 
We conjecture that a qualitatively similar bound on the strength of residual interactions, which is often stronger than Planckian, may apply to metals more generally.
\end{abstract}

\maketitle

\textbf{Introduction. } 
The electrical resistivity of conventional metals varies linearly with temperature $T$ in the regime $T \gtrsim \omega_0$, where $\omega_0$ is a characteristic phonon frequency.
The corresponding transport scattering rate extracted via Drude analysis is $1/\tau_\text{tr} = \alpha T $ (in units where $\hbar = k_B = 1$). Ambiguities associated with the Drude fit notwithstanding, it was observed that across a wide range of materials, the values of the dimensionless constant $\alpha$ are bounded by a number of order one~\cite{bruin2013similarity}. 
In conventional Migdal-Eliashberg-Bloch-Gr\"uneisen (MEBG) theory, $\alpha = 2 \pi\lambda$, where $\lambda$ is a suitably defined dimensionless electron-phonon coupling constant which is not a-priori bounded.
The observed bound is therefore striking, and has stimulated considerable theoretical activity, especially insofar as it coincides with a possible bound (christened ``Planckian''~\cite{zaanen2004temperature}) on local equilibration rates in unconventional materials such as the cuprates~\cite{hartnollmackenzie}.
An attractive feature of this idea is that it might transcend any quasiparticle-based theoretical framework, and hence give insight into a set of puzzling phenomena which have been variously identified as ``bad metals''~\cite{emerysak,hussey2004universality}, ``strange metals''~\cite{zaanen2019planckian,hartnollPlank}, ``marginal Fermi liquids''~\cite{varma2020colloquium,varma}, etc.

\begin{figure*}
	\centering
	\subfloat{%
	    \includegraphics[width=\columnwidth]{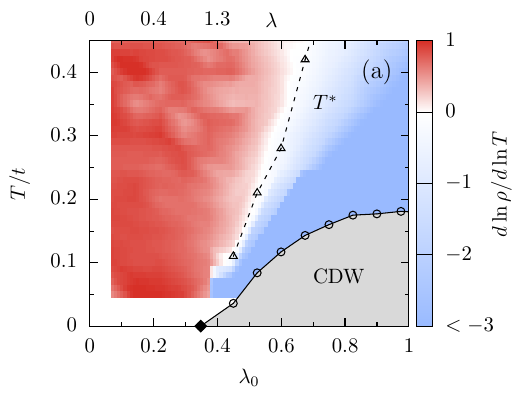}%
	}\hfill
	\subfloat{%
	    \includegraphics[width=\columnwidth]{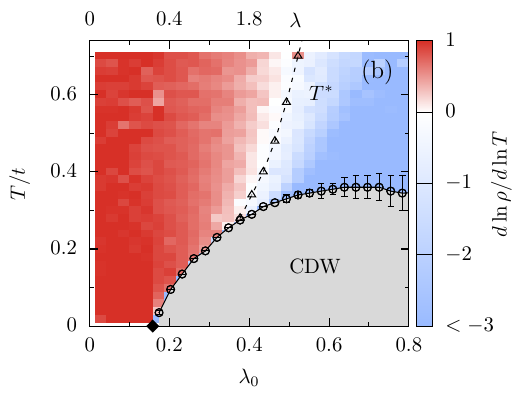}%
	}
	\caption{
	Transport phase diagram of the 2D (panel a) and 3D (panel b) Holstein model with static phonons as a function of bare (renormalized) electron-phonon coupling $\lambda_0$ ($\lambda$) and temperature $T$.
	The color scale indicates the $T$ dependence of the resistivity $\rho$, represented as an effective thermal exponent.
	CDW denotes a $(\pi,\pi)$ or $(\pi,\pi,\pi)$ charge density wave insulator.
	The $T^*$ line is a crossover at which a pseudogap in the single-particle density of states first appears, which also approximately coincides with the crossover from a ``metallic'' to ``insulating'' $T$ dependence of the resistivity. 
	The value of the renormalized coupling $\lambda$ (shown on the upper scale of the figure), which is  temperature-dependent, is computed at $T=0.25t$ in 2D and $T=0.4t$ in 3D.
	Note that in the deep blue region, the dependence on $T$ is stronger than $T^{-3}$. 
	In panel a, the chemical potential for each $\lambda_0$ is such that $n(T=0.25t) = 0.8$, while in panel b the density is $n=1$ throughout the phase diagram. 
    The calculations were done with nonzero next-nearest-neighbor hopping $t'$, with $t'=-0.3t$ in 2D and $t'=-0.2t$ in 3D. 
	}  
	\label{fig:phase_diagram}
\end{figure*}

\begin{figure*}
	\centering
	\subfloat{%
	    \includegraphics[width=\columnwidth]{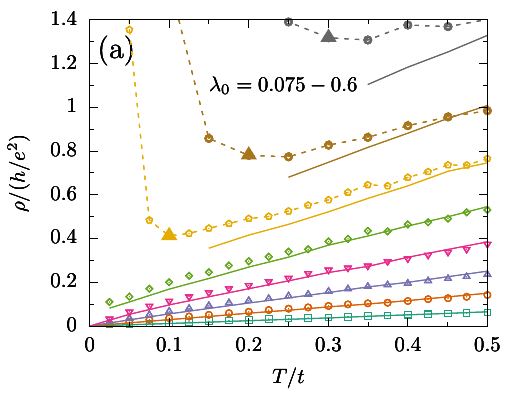}%
	}\hfill
	\subfloat{%
	    \includegraphics[width=\columnwidth]{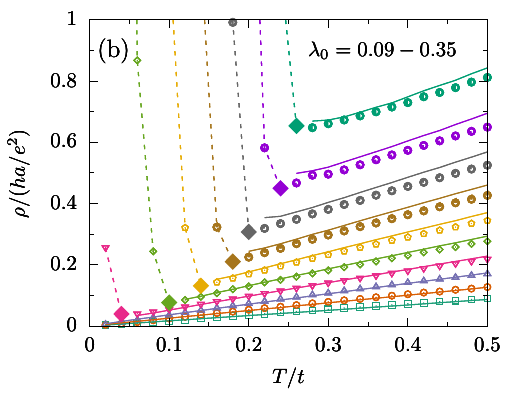}%
	}
	\caption{
    Resistivity of the 2D (panel a) and 3D (panel b) model versus temperature $T$ for 
	values of the electron-phonon coupling $\lambda_0 = 0.075$--$0.6$ in steps of $0.075$ in 2D, and $\lambda_0 = 0.087$--$0.348$ in steps of $0.029$ in 3D. 
    In the normal metallic (small $\lambda_0$) regime, the resistivity is approximately $T$-linear with zero intercept and a slope that increases (faster than linearly) with $\lambda_0$. 
    The high-temperature behavior in the strongly coupled ($\lambda_0 \gtrsim 0.4$ in 2D, $\lambda_0 \gtrsim 0.3$ in 3D) ``bad-metallic'' regime involves a resistivity that is comparable to or larger than the quantum of resistivity, and although it grows approximately linearly with $T$, it extrapolates to an increasingly large $T\to 0$ intercept, despite the absence of quenched disorder. 
	Solid lines are the Bloch-Gr\"{u}neisen (BG) formula using the renormalized $\lambda$, which is obtained from the phonon Green's function measured in Monte Carlo. 
    Dashed curves are guides to the eye in those regimes where the BG formula is not a good fit to the data. 
    In panel (a), solid triangles indicate  the resistivity at the psuedogap temperature, $T^*$. 
    In panel (b), solid diamonds indicate the same just above the CDW transition temperature.
	}
	\label{fig:rhoTlin}
\end{figure*}

We propose that, in the relevant temperature regime in metals with strong electron-phonon scattering, there is in fact generically a crossover at $\lambda \sim 1$ from metallic to insulating transport, driven by polaron physics. 
This corresponds to a bound on the slope of the $T$-linear resistivity---if $\lambda$ were any larger the system would no longer be metallic. 
Our picture comes from Monte Carlo studies of the paradigmatic Holstein model in the limit of zero phonon frequency, $\omega_0 =0$, and  more limited previous results on the breakdown of MEBG theory for $0< \omega_0 \ll T$~\cite{Esterlis:2018,Esterlis:2019,chubukov2020eliashberg,bauer2011reliability,alexandrov2001breakdown} (for a comprehensive review, see Ref.~\cite{chubukov2020eliashberg}).
The results are summarized through a phase diagram in the $\lambda$-$T$ plane in Fig.~\ref{fig:phase_diagram} and resistivity curves at various $\lambda$ in Fig.~\ref{fig:rhoTlin}.
While the proposed stability bound on $\lambda$ implies a bound on $1/\tau_{\mathrm{tr}}$ that has the same functional form as the conjectured Planckian bound, the physical origin is entirely different. 
Because scattering here is entirely elastic, the notion of a bound on thermalization of the electron fluid is irrelevant, whatever its meaning in less well-understood highly correlated materials.%
\footnote{
Electron-phonon scattering is quasi-elastic at $T \gg \omega_0$, and entirely elastic in the limit $\omega_0 \to 0$.
The fact that the stability bound on $\lambda$ holds even in this (unphysical) limit emphasizes that it is conceptually unrelated to any bound on thermalization.}

The Holstein model is at best a caricature of any actual metal, and has no direct relevance to more complicated problems in which electron-electron interactions play a central role. 
Nonetheless, we conjecture that the inferred stability bound is broadly relevant in real materials, with the caveat that the precise value of $\alpha$ at the crossover point beyond which metallic behavior ceases depends on microscopic details.
This conjecture rationalizes the otherwise surprising observation that when measured values of $\lambda$ are tabulated in conventional metals, no values larger than $\lambda \approx 2$ are found~\cite{allen,allendynes}.
Extending this intuition to more general (and less well understood) problems, we further conjecture that the coefficient $\alpha$ in any metallic system exhibiting a $T$-linear resistivity can intuitively be associated with the strength of interactions among its low-energy degrees of freedom, and so a bound on $\alpha$ reflects a bound on this interaction strength consistent with the existence of the metallic state.

\vspace{0.5em}
\textbf{Results for the Holstein model. }
We consider the Holstein Hamiltonian describing one band of non-interacting electrons coupled to an Einstein phonon, 
\begin{equation}
\label{eq:Holstein_Hamiltonian}
	H = \sum_{ij \sigma} t_{ij} c^\dag_{i\sigma} c_{j\sigma} + \sum_i \left(\frac{p_i^2}{2M} + \frac 12 K x_i^2\right)  + \gamma \sum_{i\sigma} x_i n_{i\sigma} ,
\end{equation}
where $c^\dag_{i\sigma}$ creates an electron on site $i$ with spin $\sigma$, $t_{ij}$ is the hopping matrix element between sites $i$ and $j$, $x_i$/$p_i$ are the phonon displacement/momentum on site $i$, $\omega_0 = \sqrt{K/M}$ is the phonon frequency, and $\gamma$ is the electron-phonon coupling constant, which couples the oscillator displacement to the total electron density on site $i$.
%
%
The important dimensionless parameters are the coupling strength, conventionally defined as $\lambda_0 = \gamma^2 N_0/K$, and the retardation parameter $\omega_0/E_F$, where $N_0$ is the bare density of states at the Fermi energy, $E_F$. 
It is important to distinguish between the bare coupling, $\lambda_0$, and the renormalized coupling, $\lambda$. 
The latter is the more physically relevant quantity and is defined in terms of an appropriate average of the inverse of the renormalized stiffness, $\widetilde K(\mathbf q)$, at wavevector $\mathbf q$. 
There are in fact different definitions of $\lambda$, corresponding to different averages over $\mathbf q$. 
In our studies we find the different commonly used averages give essentially the same value. 
Common definitions of $\lambda$ are summarized in 
Appendix~\ref{app:lambda}.

With some important exceptions, the regime of the Holstein model relevant to conventional metals is $\omega_0/E_F \ll 1$.
A representative phase diagram of the model in this limit as a function of $\lambda$ and $T$ is shown in Fig.~\ref{fig:phase_diagram}.
At low temperatures, there is generically a superconducting phase for weak to moderate coupling when $T < T_\text{SC} \sim \omega_0 e^{-1/\lambda}$. 
(The superconductor does not appear in our phase diagram because we will consider the limit $\omega_0 \to 0$.) 
There is an insulating charge-density wave (CDW) phase for stronger coupling when $T < T_\text{CDW}$,%
\footnote{Our CDW transition temperatures are in agreement with those reported for the 2D and 3D Holstein model with finite $\omega_0$.~\cite{Cohen-Stead:2020}}
where for large $\lambda$ (not shown in the figure) $T_\text{CDW} \sim t/\lambda$~\cite{Esterlis:2019}.
Qualitatively similar phase diagrams have been derived previously, for instance in Refs.~\cite{millis,ciuchi1999charge,fratini2021displaced}.
Here our principal interest will be in the transport properties in the disordered ``high-temperature'' regime, where $T > \omega_0$ (and hence $T > T_\text{SC}$) and $T > T_\text{CDW}$, but still $T \ll E_F$. 

When $T > \omega_0$ the phonons are effectively classical, and so we will consider a simpler version of Eq.~\eqref{eq:Holstein_Hamiltonian} in which we take $M \to \infty$, implying $\omega_0 \to 0$, and study the model via Monte Carlo simulation. 
Calculations with classical phonons are significantly simpler computationally and also allow for evaluation of dynamical observables without the need for analytic continuation. 
Moreover, we have previously~\cite{Esterlis:2018,Esterlis:2019} verified that results for various thermodynamic observables in the temperature range of interest are unchanged if calculations are carried out with finite $\omega_0$.
Dividing the Hamiltonian into phonon-only and other terms, $H = H_\text{ph} + H_\text{e}$, the thermal average of any electronic observable $\mathcal{O}$ is given, in the $M \to \infty$ limit, by
\begin{equation}
\label{eq:Annealed_Average}
    \langle \mathcal{O} \rangle \propto \int DX \, e^{-\beta H_\text{ph}[X] + \ln Z_\text{e}[X]} \, \mathcal{O}[X] ,
\end{equation}
where $Z_\text{e}[X] = \Tr e^{-\beta H_\text{e}[X]}$ is the electronic partition function and $\mathcal{O}[X] = \Tr(\mathcal{O}\,e^{-\beta H_\text{e}[X]})/Z_\text{e}[X]$ the thermal average of the observable for a given static phonon configuration $X = \{x_i\}$. 
The integral over $X$ is performed by Monte Carlo sampling. 
Further details of the algorithm are summarized in Ref.~\cite{Esterlis:2019} and in 
Appendix~\ref{app:calc}.%
\footnote{There is an extensive body of work on the problem of itinerant electrons coupled to classical spin degrees of freedom, related to the problem of magnetoresistance in manganites~\cite{Dagotto:2001}. 
While the numerical techniques are similar to those used here, the physics is distinct. 
For instance, the fixed-length constraint on the spins implies drastically different transport at elevated temperatures.}

The data presented here were computed using a two-dimensional square lattice with periodic boundary conditions and linear sizes $L \leq 20$ and a three-dimensional cubic lattice with $L \leq 14$. 
We present results for the case in which $t_{ij}$ contains nearest-neighbor hopping $t$ and next-nearest-neighbor hopping $t' = -0.3t$ (2D) or $t' = -0.2t$ (3D). 
In 2D we have fixed the chemical potential such that the average density is $n(T=0.25t)=0.8$.%
\footnote{Above the CDW transition, the density varies weakly with temperature. 
In the CDW phase, the density approaches $n=1$; see Fig.~3 of Ref.~\cite{Esterlis:2019}.}
In 3D we have fixed the average density to one electron per site, $n=1$, at all $T$. 
In the non-interacting limit, $E_F \approx 1.8 t$ (2D) and $E_F \approx 3.1 t$ (3D).
We have verified that none of the results are qualitatively sensitive to the particular choice of parameters or model details.
In 
Appendix~\ref{app:additionaldata},
we report additional data demonstrating the insensitivity of our results to varying electron density or including explicit phonon anharmonicity.

The main observable of interest is the conductivity, $\sigma(\omega)$, which refers here just to its real part. 
For a given static phonon configuration $X$, this is computed as (here $\hbar = 1$):
\begin{align}
	\sigma(\omega;X) = \frac{1}{L^d} \frac{2\pi}{\omega}\sum_{\nu \nu'} \, &[f(E_\nu) - f(E_{\nu'})] \notag \\*
	&\times |\langle \nu | \hat J | \nu' \rangle |^2 \delta(\omega - E_\nu + E_{\nu'}) ,
\end{align}
where $E_\nu$ and $|\nu \rangle$ denote single particle eigenvalues and eigenvectors of $H_\text{e}[X]$, $\hat J$ is the single-particle current operator, and $f(E) = [1 + \exp(\beta E)]^{-1}$ is the Fermi function. 
The factor of 2 accounts for spin.
This quantity is then averaged over equilibrium phonon configurations as in Eq.~\eqref{eq:Annealed_Average}. 
We denote the average simply as $\sigma(\omega)$ and the resistivity is $\rho = 1/\sigma(0)$.
There are subtleties concerning the way the dc and thermodynamic limits are taken which we discuss in detail in 
Appendix~\ref{app:subtleties}.

Our principal findings are summarized in Fig.~\ref{fig:phase_diagram}, which shows  $d\ln\rho/d\ln T$ through the phase diagram in the $(\lambda_0, T)$ plane, and in Fig.~\ref{fig:rhoTlin}, which plots $\rho(T)$ versus $T$ for various $\lambda_0$. 
The corresponding values of the renormalized coupling $\lambda$ are also reported in the figures.

Clearly, for $\lambda_0 \gtrsim 0.5$ the low-temperature CDW (a true broken-symmetry insulator) melts to a state with finite but insulating resistivity, to wit $d\rho/dT < 0$. 
Above a temperature $T^*_\rho$ which, for large $\lambda_0$, is roughly the bipolaron binding energy, $T^*_\rho \approx \gamma^2/K$, we find that $d\rho/dT > 0$ again, but with a substantial non-zero extrapolated $T\to 0$ intercept despite the absence of disorder.
$T^*_\rho$ approximately coincides with the appearance of a pseudogap in the single-particle density of states, indicated by $T^*$ in our phase diagram~\cite{Esterlis:2018,Esterlis:2019}.
The essential observation is that for the range of temperatures we are interested in ($\omega_0 \ll T \ll E_F$) there is a sharp metal-to-insulator crossover in the resistivity at $\lambda \sim 1$ driven by pseudogap formation, and hence a bound on the metallic $T$-linear resistivity.

We stress that this metal-to-insulator crossover is a reflection of local polaron physics, and is conceptually unrelated to a low-temperature CDW transition. 
Where $T^*$ is well above $T_\text{CDW}$, the CDW correlation length is small at the crossover. 
We have also carried out simulations at lower densities where charge ordering is suppressed, yet we find $T^*$ remains essentially unchanged (see 
Appendix~\ref{app:additionaldata}).
In addition, we can accurately describe the crossover using an approximation that completely neglects correlations between phonon displacements on different sites, as discussed below.

\begin{figure*}[t!]
	\begin{center}
	\includegraphics[width=\textwidth]{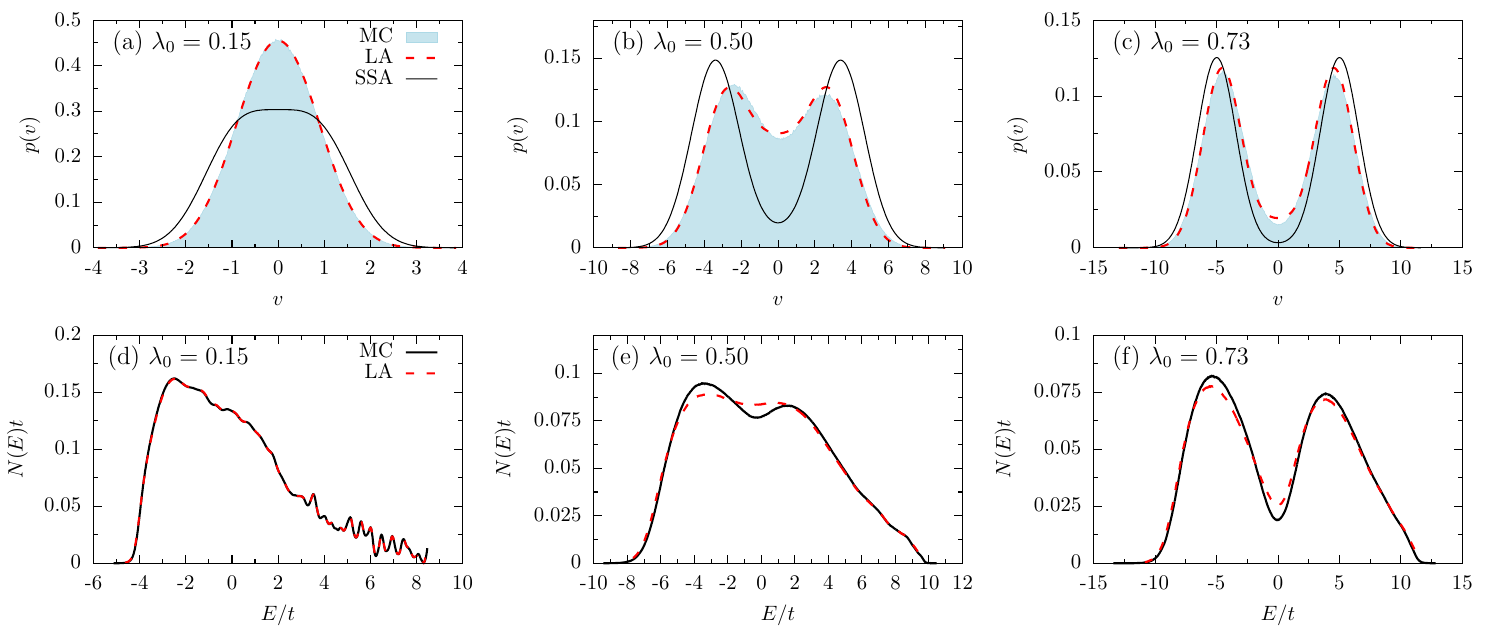}
	\end{center}
	\caption{
    Distribution of on-site potentials (panels a--c) and electronic density of states (panels d--e) in 3D, for values of $\lambda_0$ on the metallic side of the phase diagram (a \& d), near the psuedogap crossover (b \& e), and in the insulating regime (c \& f). 
    The temperature is $T=0.5t$. 
    In panels a--c, we show the on-site distributions from Monte Carlo (MC), a fit to the MC using the local approximation (LA), and the single site approximation (SSA) (see text for more details). 
    The measured distributions are accurately captured by the LA, while the SSA is only qualitatively accurate. 
    Panels d--f show the density of states as measured in MC and that obtained from a quenched disorder average using the LA. 
	Data are for $L=10$; the wiggles in the density of states (especially in d) are finite size effects that become less pronounced with increasing $L$.
	}
	\label{fig:distb+dos}
\end{figure*}

\vspace{0.5em}
\textbf{Relation to theory. } 
On the metallic side of the phase diagram we find that the conventional MEBG theory captures remarkably well the behavior of thermodynamic observables~\cite{Esterlis:2019} as well as the dc resistivity.
Comparisons of the resistivity between the MEBG theory and Monte Carlo are shown in Fig.~\ref{fig:rhoTlin}.
(The agreement with MEBG theory in the metallic regime further validates the use of $M \to \infty$ in the Monte Carlo simulations.)
A different approach is required for the insulator and crossover region. 
We have found that modelling the phonons in a ``local'' approximation as disorder with vanishing correlation length---i.e.~ignoring correlations between phonon displacements on different sites---produces reasonable quantitative agreement with the Monte Carlo data.

Formally, the Holstein model with $\omega_0 = 0$ constitutes an annealed disorder problem; see Eq.~\eqref{eq:Annealed_Average}.
The joint probability distribution for the site potentials $v_i=\gamma x_i$ is $P[V] \propto e^{-\beta H_\text{ph}[V] + \ln Z_\text{e}[V]}$, where $Z_\text{e}[V] = \Tr e^{-\beta H_\text{e}[V]}$ is the electronic partition function computed for the given potential realization $V = \{v_i\}$.
In general $P[V]$ is a complicated non-local $T$-dependent object.
We would like to approximately replace it with a local disorder distribution $P_\text{loc}[V] = \prod_i p(v_i)$.
The on-site distributions $p(v_i)$, in turn, can be extracted from the Monte Carlo data, with representative results shown in Fig.~\ref{fig:distb+dos}.

We can obtain a crude representation of $p(v)$ by considering the statistical properties of a single isolated site (no hopping), for which we can compute the phonon distribution $p_\text{ss}(v)$ exactly as a function of the chemical potential $\mu$ and temperature $T$:
\begin{equation}
\label{eq:single_site_distbn}
p_\text{ss}(v) = p_0 \! \left[
    e^{-\frac{(v-U_0)^2}{2UT}}
    + 2y e^{-\frac{v^2}{2UT} -\frac{U_1}{2T}}
    + y^2 e^{-\frac{(v+U_0)^2}{2UT}} \right] ,
\end{equation}
where $U_0 = U_1 = U \equiv \gamma^2/K$ is the characteristic bipolaron binding energy, $y=e^{\mu/T}$ is the electron fugacity ($y=1$ when there is on average one electron per site), and $p_0$ is the requisite normalization factor.
This approximation becomes better the deeper we are in the insulating phase, i.e.~the more localized the electronic states in a typical realization of $V$ are. 
In the intermediate range of $\lambda$ of primary interest here, $p(v)$ deviates substantially from $p_\text{ss}(v)$, but it can be well parameterized by the same functional form with $U_0$ and $U_1$ treated as $T$ and $\lambda$ dependent parameters; we therefore do this when making direct comparisons with the numerical data.
One can then calculate electronic observables in realizations of $V$ and perform a quenched average using $P_\text{loc}[V]$.

Roughly speaking, $p_\text{ss}(v)$ contains peaks at $v=\pm U$, each of width $\sqrt{TU}$.
At strong coupling ($\lambda_0 > 0.5$), when $U$ is larger than both the unperturbed bandwidth and $T$, the disorder distribution is bimodal, and the density of states itself splits into two peaks, such that the integrated density of states per spin in the low-energy peak is $n/2$.%
\footnote{In strong coupling and ignoring thermal broadening, the effective disorder distribution is that of a binary alloy---with energy $-U$ on the ``bipolaron sites'' and energy $+U$ on the remaining sites.  
Manifestly, the concentration of bipolaron sites is $n/2$.  
When $2U$ is larger than the bandwidth, by the spectral localization theorem~\cite{ehrenreich1976} this would give rise to a hard gap and an integrated density of states per spin in the lower band precisely equal to $n/2$.  
Thermal broadening turns this gap into a pseudo-gap, which survives even when $2U$ is somewhat less than the bandwidth.}
The chemical potential thus automatically lies in the pseudogap between them---this contrasts with the case of a corresponding problem with quenched disorder where the chemical potential is an independent quantity that is not generically tied to the pseudogap.
This band splitting captures the binding of electrons into bipolarons. 
The states deep in the pseudogap are strongly localized. 

However, even at strong coupling, increasing $T$ beyond $U$ makes the disorder distribution single-peaked, and there is no pseudogap.
Although the effective disorder is relatively strong here, one can crudely understand the observed $T$ dependence of the resistivity with perturbative reasoning---assuming that the scattering rate (and hence the resistivity itself) is proportional to the mean-square disorder potential, $\overline{(v_i - \bar{v})^2} \sim U^2 +UT$.
This accounts both for the observed linear-in-$T$ growth and the large extrapolated $T \to 0$ intercept of $\rho$ in Fig.~\ref{fig:rhoTlin} in this range of $T$ and $\lambda_0$.

Ultimately, the observed metal-insulator crossover apparent in the Monte Carlo data occurs for two closely intertwined reasons: the band splitting causes a depression in the density of states (pseudogap) at the Fermi level, and these states become more and more localized. 
The single site approximation captures qualitatively the essential physics of the crossover, and is quantitatively accurate at sufficiently strong coupling. 
Over a much broader range of couplings, we are still able to fit the distribution of site energies measured in the Monte Carlo data by treating $U_0$ and $U_1$ in Eq.~\eqref{eq:single_site_distbn} as adjustable parameters.
Observables computed using the resulting $P_\text{loc}[V]$ agree reasonably well with their Monte Carlo values throughout the phase diagram (except, of course, in or very near the CDW phase).
Representative results are shown in Fig.~\ref{fig:distb+dos}.

Note that for large $\lambda$, the local approximation is equivalent to previous results obtained using dynamical mean field theory (DMFT) for the same problem~\cite{freericks1993,millis,ciuchi1999charge}.  
Indeed, the DMFT results appear to agree---at least qualitatively---with our Monte Carlo results, even at small $\lambda$ where the naive single-site approximation (with $U_0 = U_1 = U$) fails.

\vspace{0.5em}
\textbf{Stability bounds. } 
While the results obtained concern a simplified model, we expect the qualitative and even the rough quantitative aspects of the results to apply in realistic circumstances in which electron-phonon coupling is strong.  
There are a number of material systems which, as a function of pressure, strain, doping concentration, or even photo-excitation, can be tuned from a metallic or superconducting to an insulating CDW ground state.  
To the extent that these low temperature phase transitions reflect changes in the strength of the electron-phonon coupling $\lambda$, a corresponding crossover should be expected at elevated temperature in the ``normal state'' from metallic $T$ dependence on the small $\lambda$ side of the crossover to increasingly insulating $T$ dependence on the large $\lambda$ side.  
Moreover, at the crossover, the scattering rate should be $1/\tau \sim 2\pi T$ (corresponding to $\lambda \sim 1$).  
A systematic comparative study of such crossovers in a variety of materials would be an interesting way to test the relevance of the present studies.  
Examples of systems which at least superficially exhibit some aspects of this expected behavior include certain organics~\cite{ito1996metal,limelette2003mott} and BPBO~\cite{uchida1987superconductivity}.

More generally, the appeal of invoking a Planckian bound $1/\tau_\text{eq} \leq 2 \pi T$ in relation to transport is the hope that it gives a way to understand the origin of $T$-linear resistivity in a variety of unconventional metallic systems.
A priori, the above discussion is applicable only to conventional metals, where the physics of the $T$-linear resistivity has been well understood for decades and the only new insight we have to offer concerns why larger values of $\alpha$ are never observed. 
However, one can speculate that there is a broader sense in which the present results may inform the discussion of less well understood metallic systems as well~\cite{cooper2009anomalous}.

It is certainly conceivable that a $T$-linear resistivity can arise in a semiclassical regime in which electrons scatter from  another form of collective soft mode of a system, other than a phonon. 
Here, the same considerations we have explored above apply more or less directly.

More generally, it is a common (not necessarily universal) feature of quantum systems that strong interactions among propagating particles reduce itineracy. 
Thus, it is reasonable to suppose---in the absence of disorder---that the resistivity (or more directly $1/\tau_\text{tr}$) is a measure of the strength of the residual interactions between low-energy degrees of freedom.  
At the same time, there is general sense in which strong interactions lead to a reorganization of the effective low energy degrees of freedom.  
This can be made precise in some systems in which there is an explicit transformation relating a set of interacting ``microscopic'' variables to a set of dual degrees of freedom, such that when the former is strongly interacting the latter is weakly interacting, and vice versa.  
But the underlying physical intuition is likely more broadly valid---that there is a rough maximum strength of an appropriate dimensionless measure of the effective interactions consistent with a metallic state.

Thus, we conclude with the conjecture that there is a general stability bound on the magnitude of the resistivity of any metallic state---one that often is much more restrictive (and hence more significant) than the putative Planckian bound, as we explain in 
Appendix~\ref{app:rates}. 
Where $1/\tau_\text{tr} \approx \alpha T$, even when this is not directly attributable to electron-phonon scattering, it is reasonable to suppose that $\alpha$ is the correct dimensionless measure of the interaction strength, and so is bounded by these considerations.%
\footnote{For example, in Ref.~\cite{georges}, the transport properties of the Hubbard model with intermediate $U$ were analyzed using dynamical mean field theory (DMFT), with particular focus on an intermediate temperature regime $T_\text{FL} < T < T_\text{MIR}$, where $T_\text{FL}$ is a temperature below which Fermi liquid theory applies and $T_\text{MIR}$ is the temperature above which the resistivity exceeds the quantum of resistance. 
In this regime, the resistivity is found to be approximately linear in $T$ with a slightly negative extrapolated value at $T\to 0$.  
An interpretation of the results in terms of highly dressed ``resilient quasi-particles'' is shown to account for the behavior qualitatively, where a suitably defined transport scattering rate depends on hole-doping and $T$ (and, presumably, $U/t$), but is ``at most comparable to $T$.''  
This model is conceptually unrelated to the electron-phonon problem we have analyzed, but it is plausible that a related stability bound is the explanation for the apparent bound on $1/\tau_\text{tr}$.}
Where $1/\tau_\text{tr}$ has a more complicated $T$ dependence, it requires further analysis to relate its magnitude to a dimensionless interaction strength.  
However, in some cases such a relationship can be established on other grounds.
For example, at low $T$, electron scattering from long-wavelength acoustic phonons dominates the resistivity of many metals, such that $1/\tau_\text{tr} \sim \lambda_{D} T^5/\omega_D^4$ where $\omega_D$ is the Debye frequency, which can be independently determined. 
Thus, a bound on $\lambda_D$ implies a corresponding bound on the resistivity.  
One can also consider extracting an estimate of the strength of the residual interactions from dimensional analysis, as $\lambda_\text{eff} \sim 1/\Omega\tau_\text{tr}$, where $\Omega$ is a characteristic energy scale in the problem.  
In most metals, a lower bound on $\lambda_\text{eff}$ can be obtained by taking $\Omega=E_F$, since typically $E_F$ is the largest characteristic scale. 
This leads to a somewhat different perspective on the  familiar Ioffe-Regel limit, usually stated as $E_F \tau_\text{tr} \sim k_F \ell \gtrsim 1$.
 
In summary, we propose that, while the precise way in which it plays out can vary depending on specifics, there is an approximate stability bound on the maximum magnitude of an appropriate dimensionless measure of the transport scattering rate in all clean metallic systems.%
\footnote{To avoid misunderstanding, we review the fine print on this proposal:  
The proposed bound is approximate in the sense that it involves a dimensionless number of order one that can depend on microscopic details. 
However, it appears to be difficult to find physically reasonable circumstances in which this number is substantially larger than 1.  
It is only indirectly related to a resistivity bound, in the sense that $\rho$ is proportional to $1/\tau_\text{tr}$.  
As already discussed, determining the appropriate energy scale to relate the dimensional $1/\tau_\text{tr}$ to a dimensionless coupling constant generally involves additional analysis, but in circumstances in which the scattering rate is $T$-linear, the correct dimensionless quantity is $1/\tau_\text{tr} T$.}
\\

\begin{acknowledgments}
We gratefully acknowledge conversations with Kamran Behnia, Erez Berg, Sankar Das Sarma, Luca Delacr\'etaz, Matthew Fisher, Jim Freericks, Sarang Gopalakrishnan, Tarun Grover, Nigel Hussey, Prashant Kumar, Andy Mackenzie, Chetan Nayak, Brad Ramshaw, John Sous, and especially with Sean Hartnoll.
This work was supported in part by the Gordon and Betty Moore Foundation's EPiQS Initiative through GBMF8686 (CM), the Stanford Graduate Fellowship (AP), and NSF grant No.~DMR-2000987 at Stanford (SAK).
CM also acknowledges the hospitality of the Kavli Institute for Theoretical Physics, supported by the National Science Foundation under grant No.~NSF PHY-1748958.
\end{acknowledgments}


%

\appendix

\section{Calculating the conductivity}
\label{app:calc}

\subsection{Monte Carlo algorithm}

Here we describe the algorithm used for the 3D simulations presented in the main text. Details on the techniques used for 2D may be found in Ref.~\cite{Esterlis:2019}.

In terms of the quantities $v_i = \gamma x_i$ and $U = \gamma^2/K$, the Holstein Hamiltonian at $M = \infty$ is
\begin{equation}
H-\mu N  = \frac{1}{2U} \sum_i v_i^2 + H_{\text{e}}[V],
\end{equation}
where
\begin{equation}
H_{\text{e}}[V] = \sum_{ij\sigma} t_{ij} c^\dagger_{i\sigma} c_{j\sigma} +\sum_{i\sigma} (v_i-\mu) n_{i\sigma} .
\end{equation}
The thermal average of any electronic operator $\mathcal{O}$ is then
\begin{equation}
\langle \mathcal{O} \rangle = \int DV \, P[V] \, \mathcal{O}[V] ,
\end{equation}
where $\mathcal{O}[V] = \Tr(\mathcal{O} e^{-\beta H_{\text{e}}[V]}) / Z_{\text{e}}[V]$,
\begin{equation}
P[V] = \frac{1}{Z} \exp\!\left(-\frac{\beta}{2U} \sum_i v_i^2 + \log Z_{\text{e}}[V] \right) ,
\end{equation}
and $Z_{\text{e}}[V] =  \Tr(e^{-\beta H_{\text{e}}[V]})$; the traces are over the electronic degrees of freedom.

We sample from the distribution $P[V]$ using the well-known Langevin Monte Carlo algorithm~\cite{rossky1978}.
Global configuration updates $V \to V'$ are proposed according to an effective Brownian dynamics:
\begin{equation}
v_i' = v_i - \tau F_i[V] + \sqrt{2\tau} \xi_i ,
\end{equation}
where
\begin{equation}
F_i[V] = - \frac{\partial \log P[V]}{\partial v_i} 
= \beta \left( \frac{v_i}{U} + n_i[V] \right) ,
\end{equation}
each $\xi_i$ is independently sampled from the standard normal distribution (mean 0, variance 1), and $n_i[V]$ is the expected number of particles on site $i$ in configuration $V$.
Proposed updates are then accepted with probability
\begin{equation}
p_{\text{acc}} = \min\!\left\{ 1, \, \frac{P[V'] Q[V' \to V]}{P[V] Q[V \to V']} \right\} ,
\end{equation}
where $Q[V \to V']$ is the transition probability
\begin{equation}
Q[V \to V'] \propto \exp\!\left[ - \frac{1}{4\tau} \sum_i \big( v_i' - v_i + \tau F_i[V] \big)^2 \right] .
\end{equation}
Note that $Q[V' \to V] \neq Q[V \to V']$ because $F_i[V'] \neq -F_i[V]$ in general.
We perform an acceptance/rejection step after each Langevin step.

For each given configuration $V$, we construct the single-particle Hamiltonian
\begin{equation}
\mathcal{H}_{ij}[V] = t_{ij} + (v_i-\mu) \delta_{ij} ,
\end{equation}
and diagonalize it to obtain single-particle eigenfunctions $\psi_{ia}[V]$ and eigenvalues $E_a[V]$, in terms of which
\begin{equation}
\log Z_{\text{e}}[V] = 2 \sum_a \log(1+ e^{-\beta E_a[V]})
\end{equation}
and
\begin{equation}
n_i[V] = 2 \sum_a \frac{|\psi_{ia}[V]|^2}{1 + e^{\beta E_a[V]}} .
\end{equation}
We also compute matrix elements of the single-particle current operator
\begin{equation}
\label{eq:current_matrix_elements}
\langle a | \hat{J} | b \rangle = - i e \sum_{ij} (r_i - r_j) t_{ij} \psi^*_{ia}[V] \psi_{jb}[V] ,
\end{equation}
where $r_i$ is the appropriate component of the position of lattice site $i$.

In each Monte Carlo simulation from which data is shown, we first thermalized the system for $N_\text{therm} \geq 5000$ Langevin steps, and then averaged the quantities of interest over up to $N_{\text{samp}} = 10^5$ sampled configurations. 
The effective timestep $\tau$ was chosen so that the acceptance rate of proposed updates is approximately $0.5$.
To reduce finite-size effects, we performed simulations with a small orbital magnetic field $B$ corresponding to a single flux quantum penetrating the system ($B \propto 1/L^2$ vanishes in the thermodynamic limit), implemented by multiplying all hopping integrals $t_{ij}$ in the above formulae by the appropriate Peierls phase factors. In the 3D simulations, $B$ was taken to point along the $z$ axis, and we have reported $\rho_{zz}$ (we have verified that $\rho_{xx}$ tends towards the same value as $L$ is increased, as required by the cubic symmetry of the $L \to \infty$ problem).

\subsection{Optical conductivity}

\begin{figure*}
	\begin{center}
	\includegraphics[width=\textwidth]{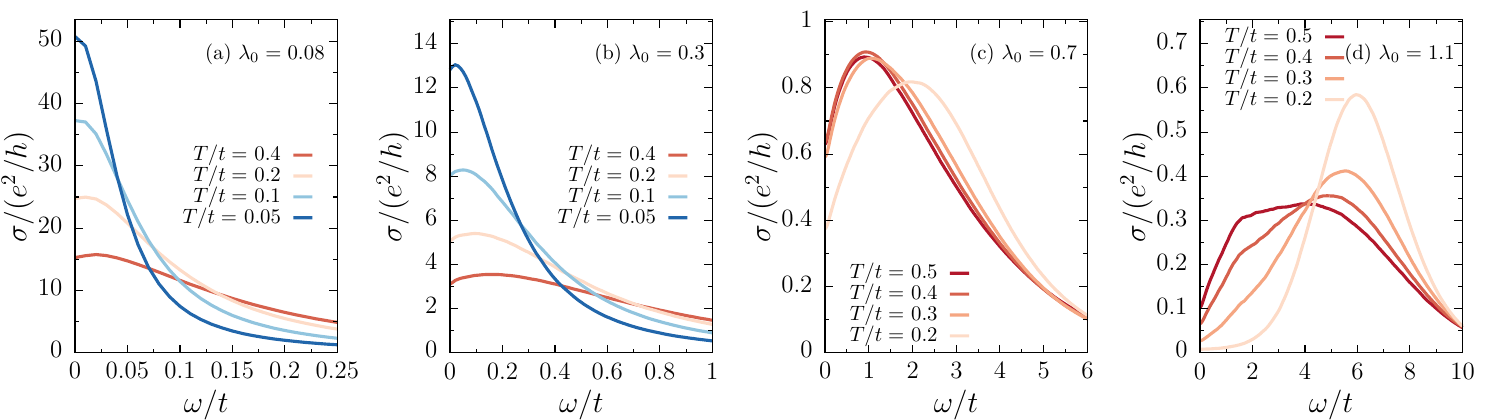}
	\end{center}
	\caption{
	Optical conductivity, $\sigma(\omega)$, in 2D for several values of the bare coupling $\lambda_0$ and temperatures, $T$. 
    Other parameters are the same as those used to generate the phase diagram in Fig.~1a of the main text. 
    Panels a and b correspond to representative $\lambda_0$ values on the metallic side of the phase diagram, while c and d are on the insulating side.
	}
	\label{fig:sigma}
\end{figure*}

Figure~\ref{fig:sigma} shows the real part of the optical conductivity for several values of the electron-phonon coupling and at several temperatures in 2D. 
We denote the real part as $\sigma(\omega)$. 
In any given phonon configuration, it is computed according to ($\hbar = 1$):
\begin{equation}
	\sigma(\omega) = \frac{2\pi}{L^d} \sum_{ab} 
	\left[\frac{f(E_a) - f(E_b)}{E_b - E_a} \right] \, 
    |\langle a | \hat{J} | b \rangle |^2 \,
	\delta(\omega - E_a + E_b) ,
\end{equation}
where $E_a$ and $|a \rangle$ denote single-particle eigenvalues and eigenvectors computed in the given phonon configuration, $\hat{J}$ is the current operator whose matrix elements are given by Eq.~\eqref{eq:current_matrix_elements}, and $f(E) = [1 + \exp(\beta E)]^{-1}$ is the Fermi function. 
To obtain a continuous result, we replace the $\delta$-function by a Lorentzian in $\omega$ of width
\begin{equation}
\label{eq:Gamma}
    \Gamma(L) = \Gamma_0 t/L^2 ,
\end{equation}
where $L$ is the linear dimension of our system;  this choice ensures that $\Gamma$ vanishes as we approach the thermodynamic limit, while for $\Gamma_0$ sufficiently large (we take $\Gamma_0=1$), this broadening is sufficient to eliminate the structure due to the discreteness of the levels in a finite-sized system.  
The result is then averaged over phonon configurations as described in the previous section, and the results are extrapolated to the $L \to \infty$ limit by fitting a straight line to $\sigma(\omega;L)$ versus $\Gamma(L)$.

Figures~\ref{fig:sigma}a,b show $\sigma(\omega)$ for two representative values of $\lambda_0$ on the metallic side of the phase diagram in 2D. The conductivity is a decreasing function of temperature and (at low $T$) exhibits a maximum at $\omega=0$.%
\footnote{In some cases, even where the system is metallic, the peak in the  optical conductivity $\sigma(\omega)$ is slightly displaced from $\omega = 0$.~\cite{hussey2004universality}  
The existence of a ``displaced Drude peak'' similar to what we observe was discussed in Ref.~\cite{fratini2021displaced}.}
Figures~\ref{fig:sigma}c,d show $\sigma(\omega)$ for two representative values of $\lambda_0$ on the insulating side, again in 2D. 
Here the conductivity is an increasing function of temperature and exhibits a minimum at $\omega=0$.

\subsection{DC resistivity}

At high temperatures, above characteristic phonon energy scales, the Bloch-Gr\"{u}neisen formula yields the following $T$-linear behavior for the resistivity ($\hbar = e = 1$):
\begin{equation}
	\rho = \frac{2\pi \lambda_\text{tr} T}{(n/m)_\text{eff}} ,
	\label{eq:BG_formula}
\end{equation}
where
\begin{equation}
	\left(\frac{n}{m}\right)_\text{eff} = \frac{{2}}{L^d} \sum_\mathbf k v_{\mathbf{k}, x}^2\left(-\frac{\partial f}{\partial \epsilon_\mathbf k}\right),
\end{equation}
$v_{\mathbf{k}, x} = \partial \epsilon_{\mathbf{k}} / \partial k_x$,
and $\lambda_\text{tr}$ is a renormalized electron-phonon coupling constant, defined in the following section.
Note that both $(n/m)_\text{eff}$ and $\lambda_\text{tr}$ are in general (weakly) $T$ dependent, so that the $T$-linearity in Eq.~\eqref{eq:BG_formula} is only approximate.

\section{Definitions of $\lambda$}\label{app:lambda}

In the Holstein model it is always possible to define a dimensionless ratio of the bare parameters to characterize the strength of the electron-phonon coupling. 
The ``polaron'' energy is unambiguously given by $\gamma^2/K$. 
Comparing this to a characteristic electronic energy can be done in multiple ways: 
in the weak coupling limit, where all that matters are states near the Fermi energy, the relevant energy scale is the inverse of the (unrenormalized) density of states at the Fermi energy, $N_0=N(E_F)$.  
For strong coupling, the more relevant energy is the band-width, $W = 4dt$.  
Since for generic band-structures, it is typically the case that $N_0 W \sim 1$, the distinction between these two ways of defining $\lambda_0$ is not significant.  
There are special band-structures for which $N_0 W$ is either much larger than one (e.g.~at a sufficiently singular van-Hove point that $N_0$ diverges) or much smaller (e.g.~in a semi-metal with the Fermi energy at the Dirac point);  in such cases, the differences in definition are significant. 
As we have focused our discussion on generic bandstructures for metals, for ease of discussion we have simply adopted a uniform convention such that
\begin{equation}
\lambda_0 \equiv \frac{\gamma^2 N_0}{K} .
\end{equation}

Depending on the context, different definitions of a ``renormalized" coupling may be appropriate. 
In general, the renormalized coupling takes the form of a wavevector average of the inverse renormalized phonon stiffness, $\widetilde K(\mathbf{q})$:
\begin{equation}
    \lambda_w = \gamma^2 N_0 \left\langle \frac{1}{\widetilde K(\mathbf{q})} \right\rangle_w ,
\end{equation}
where the average is defined with respect to a ``weight function'' $w$ according to
\begin{equation}
\label{eq:K_avg}
      \left\langle F(\mathbf{q}) \right\rangle_w =  \frac{\sum_{\mathbf{k} \mathbf{k}'} F(\mathbf{k}-\mathbf{k}')w(\mathbf{k},\mathbf{k}') \delta(\epsilon_\mathbf{k} - E_F)\delta(\epsilon_{\mathbf{k}'}- E_F)}{\sum_{\mathbf{k} \mathbf{k}'} w(\mathbf{k},\mathbf{k}') \delta(\epsilon_\mathbf{k} - E_F)\delta(\epsilon_{\mathbf{k}'}- E_F)} .
\end{equation}
Different applications use different $w$ functions. 
In superconductivity theory, one defines $\lambda_\text{sc}$ simply with $w = 1$.  
For transport, the coupling $\lambda_\text{tr}$ is defined with respect to the weight function
\begin{equation}
    w(\mathbf{k},\mathbf{k}') = (v_{\mathbf{k},x} - v_{\mathbf{k}'\!,x})^2,
\end{equation}
where $v_{\mathbf{k}, x} = \partial \epsilon_{\mathbf{k}} / \partial k_x$.
In the main text we have used $\lambda$ to refer to $\lambda_\text{tr}$. 
However, in typical cases~\cite{allen} there is not a significant difference between $\lambda_\text{tr}$ and, e.g,~$\lambda_\text{sc}$. 
In practice, the delta functions appearing in Eq.~\eqref{eq:K_avg} have been regularized according to
\begin{equation}
    \delta(\epsilon) \to -\frac{\partial f}{\partial \epsilon},
\end{equation}
where $f(\epsilon)$ is the Fermi function.

\section{Subtleties concerning the resistivity} 
\label{app:subtleties}

Dynamical properties are notoriously subtler than thermodynamic ones and this is doubly true of the DC resistivity.  

It is conceptually obvious---and was verified explicitly in Ref.~\cite{Esterlis:2019}---that when $T > \omega_0$, thermodynamic quantities are all essentially independent of $M$, i.e.~the limit $\omega_0\to 0$ is without subtlety.  
Likewise it is clear that if we consider small perturbations to the model, such as weak coupling to acoustic phonons, that this will produce comparably weak (perturbative) corrections to thermodynamic correlations.

However, this issue is much more subtle in the case of dynamical correlations.  
Possible problems with the $M \to \infty$ limit arise because non-zero $\omega_0$ can have dramatic implications at long times ($\omega_0t \gg 1$), even if $T \gg \omega_0$.  
Qualitative effects of weak coupling to additional low-energy degrees of freedom are well known in the theory of weak localization, where the DC resistivity is found to have a singular dependence on a ``phase-breaking'' coherence length that diverges as $T \to 0$,~\cite{AAK1, AAK2} as we discuss briefly in 
Appendix~\ref{app:rates}.

\begin{figure}
	\begin{center}
	\includegraphics[width=\columnwidth]{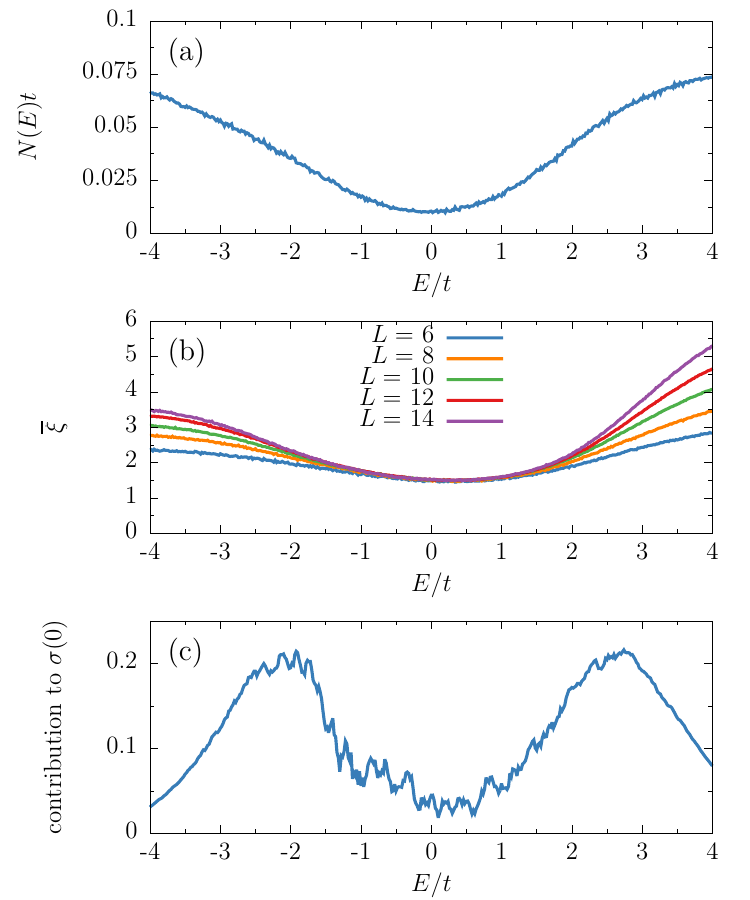}
	\end{center}
	\caption{
	Energy-resolved quantities in 3D at $\lambda_0 = 0.8$ and $T/t = 0.5$, corresponding to a point in the insulating pseudogap region of the phase diagram in Fig.~1b of the main text.
    Panel a shows the electronic density of states, $N(E)$, and panel b shows the averaged participation length, $\xi$. 
    The participation length of a normalized single-particle state $\psi$ is defined as $\xi(\psi) = (\sum_i |\psi_i|^4)^{-1/d}$, where the sum is over lattice sites $i$, and $d=3$ is the dimension of space. 
    $\overline{\xi}(E)$ is the Monte Carlo average of $\xi(\psi_a)$ over single-particle eigenfunctions $\psi_a$ whose energy $E_a \approx E$. 
    Panel c shows the relative contribution of single-particle eigenstates with energy $E$ to the DC conductivity, $\sigma(0)$. 
    States within the pseudogap are strongly localized, and the dominant contribution to $\sigma(0)$ comes from thermally excited delocalized states above an effective mobility edge. 
    Here $N(E)$ is shown for $L=12$, and $\sigma(0)$ is computed as described in 
    Appendix~\ref{app:calc}.
	}
	\label{fig:ipr_ins}
\end{figure}

\begin{figure}
	\begin{center}
	\includegraphics[width=\columnwidth]{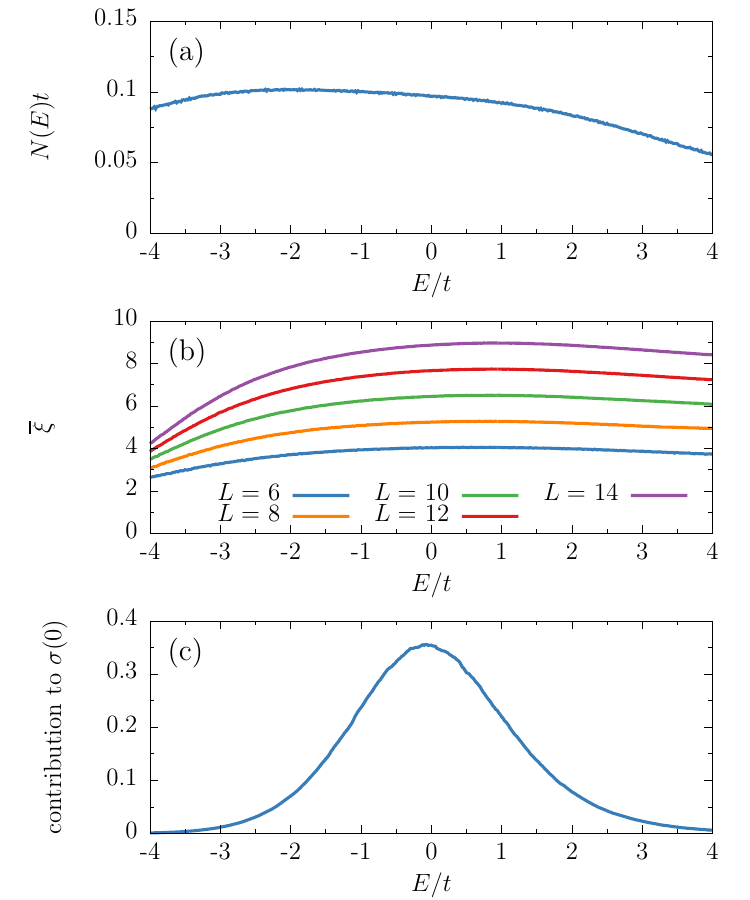}
	\end{center}
	\caption{
	Energy-resolved quantities in 3D at $\lambda_0 = 0.4$ and $T/t = 0.74$, corresponding to a point just on the metallic side of the phase diagram and at the highest temperature shown in Fig.~1b of the main text. 
    Panel a shows the electronic density of states, $N(E)$, and panel b shows the averaged participation length, $\xi$, as defined in the caption of Fig.~\ref{fig:ipr_ins}. 
    Panel c shows the relative contribution of single-particle eigenstates with energy $E$ to the DC conductivity, $\sigma(0)$. 
    Clearly, there is no pseudogap for these parameters, and $\overline{\xi}$ increases with system size $L$, indicating that states near the chemical potential are delocalized. 
    $N(E)$ is shown for $L=12$.
	}
	\label{fig:ipr_metal}
\end{figure}

Turning to the present problem, the most obvious issue is indeed localization: 
since for $\omega_0=0$, what we are dealing with is equivalent to a problem of non-interacting electrons in a random potential, all states are strictly localized in 2D. 
The same is true in 3D for large enough $\lambda_0$ and/or high enough $T$.  
Then the DC resistivity must diverge in the thermodynamic limit, $L\to \infty$. 
For the most part, we are indifferent to this since, in the parameter regime of interest, the states that dominate the resistivity are either delocalized (in the 3D case; see Figs.~\ref{fig:ipr_ins} and~\ref{fig:ipr_metal}) or have a localization length longer than the sizes of the systems we have studied.  
Our protocol for broadening the $\delta$-functions in the expression for $\sigma(\omega)$ means that our results are effectively equivalent to results in the thermodynamic limit in the presence of a phase-breaking length $\ell_\phi \sim L$. 
This can be thought of either as incorporating the physics of non-zero $\omega_0$, or as reflecting the presence of other, more weakly coupled phonon modes, including acoustic phonons.

Since, where the weak localization theory applies, the resistivity depends only logarithmically on $\ell_\phi$, the results we obtain are not expected to be strongly dependent on precisely what values we assume for $\ell_\phi$.  
We have confirmed this by testing that the present results are not extremely sensitive to the assumed value of $\Gamma_0$ in Eq.~\eqref{eq:Gamma}.

Deep in the pseudogap regime, as can be seen in Fig.~\ref{fig:ipr_ins}, the states near the Fermi energy are strongly localized but the states above a roughly defined mobility edge are effectively delocalized, as evidenced by the fact that the ``participation length'', defined as  $\xi = ( \sum_i |\psi_i|^4 )^{-1/d}$, grows (more or less linearly) with increasing $L$.
(At the very least, this implies that these high energy states are not localized on the scales of $L$ we have explored).
In this regime, the DC resistivity as we compute it is dominated by thermally excited states above the effective mobility edge.
At low enough $T$, so long as the pseudo-gap does not turn into a full gap, one would expect some interesting form of hopping conduction involving states near the Fermi energy to dominate the transport.%
\footnote{High temperature polaron hopping transport in the Holstein model was recently studied in Ref.~\cite{szabo2021}.}
However, in the relevant range of $\lambda_0$, such a regime may not exist for $T> T_{\text{CDW}}$, the temperature at which CDW order onsets.  
In any case, the strong divergence of $\rho$ with decreasing $T$ is not in question.

One may also wonder about the role of localized states at temperatures $T > T^*$ with moderately large $\lambda_0 \sim 1$ where a $T$-linear resistivity with a substantial extrapolated $T\to 0$ offset is observed, with $\rho$ comparable to or larger than the quantum of resistance. 
However, even here, in the range of $T$ we have explored, from measurements of the participation lengths one can conclude that the dominant current carrying states are delocalized (in the 3D case; see Fig.~\ref{fig:ipr_metal}) or localized on length scales that are larger than or comparable to the accessible system sizes.  
Thus it is reasonable to invoke some weak form of phase breaking as justification for accepting the results of our calculation as being physically reasonable.

\section{The difference between $\tau_\text{tr}$ and the equilibration time $\tau_\text{eq}$.}\label{app:rates}

It has been correctly pointed out~\cite{hartnollmackenzie, poniatowski2021counterexample} that $1/\tau_\text{tr}$ is {\em not} directly a measure of the rate at which a system reaches local equilibrium, i.e.~it is not the relaxation rate $1/\tau_\text{eq}$ that is subject to a putative Planckian bound.  
The clearest illustration of this is the case in which $\tau_\text{tr}$ reflects purely elastic scattering, e.g.~scattering off disorder, which provides no mechanism for the energy of the system to relax and no route for entropy production. 

The equilibration rate $1/\tau_\text{eq}$ is defined as the rate at which the system achieves local equilibrium;  it is therefore determined by the slowest relevant relaxation process.  
We expect the rate of energy relaxation of the electrons to be less than or equal to the transport decay rate, $1/\tau_\text{E} \leq 1/\tau_\text{tr}$, and in turn $1/\tau_\text{eq} \leq 1/\tau_\text{E}$.
It is thus possible to find circumstances in which $1/\tau_\text{tr} \approx 2\pi T$ and yet $1/\tau_\text{eq} \ll 2\pi T$. 
This is the case in our results for $\lambda \approx 1$. 
When $ \omega_0 \ll T$, the relevant scattering processes are quasi-elastic. 
Each electron-phonon collision exchanges energy of order $\omega_0$, so that the change in energy relative to $E\sim T$ after time $t$ is $\Delta E(t) \sim \omega_0 \sqrt{t/\tau_\text{tr}}$.%
\footnote{More precisely, $1/\tau_\text{tr}$ in this expression should be replaced with the quasiparticle scattering rate, but in present circumstances this distinction is unimportant.}
Electron thermalization occurs when $\Delta E(\tau_\text{E} )\sim E$, so
\begin{equation}
1/\tau_\text{E} \sim ( \omega_0/T)^2 \  1/\tau_\text{tr} \, .
\end{equation}
Thus, at $T \gg \omega_0$, even when $1/\tau_\text{tr} \approx 2\pi T $, $1/\tau_\text{eq}$ is parametrically smaller than the Planckian upper bound.

Also of relevance is the electron phase-breaking rate $1/\tau_{\phi}$,~\cite{AAK1, AAK2} related to the phase-breaking length $\ell_\phi$ discussed in 
Appendix~\ref{app:subtleties}.
In common with $\tau_\text{eq}$, for the present model this time scale diverges as $\omega_0 \to 0$.  
So long as the single particle states are at most weakly localized, this rate reflects the accumulation of a random phase due to energy exchange with phonons: $\Delta E(\tau_{\phi}) \tau_{\phi} \sim 1$, which gives 
\begin{equation}
\tau_{\phi} \sim \tau_\text{tr}(\omega_0 \tau_\text{tr})^{-2/3} \ll \tau_\text{E} \, ,
\end{equation}
i.e.~phase relaxes much faster than energy for small $\omega_0$. 
This timescale cuts off weak localization corrections to the classical conductivity in two dimensions (ignoring numerical coefficients):
\begin{equation}
    \sigma \sim \frac{n e^2 \tau_\text{tr}}{m} - \frac{e^2}{h}\ln \frac{\tau_{\phi}}{\tau_\text{tr}} 
    \sim \frac{e^2}{h} \left[ E_F\tau_\text{tr}+ \ln(\omega_0\tau_\text{tr}) \right] \, .
\end{equation}
Given the logarithmic dependence on $\omega_0$, this contribution can at least be neglected in the robustly metallic state when $\sigma \gg  e^2/h$. 
On the other hand, as has been emphasized in Ref.~\cite{fratini2021displaced}, weak localization may be an essential part of the physics (even at small $\lambda$ and non-zero $\omega_0$) when the resistivity is large enough, in particular in the high-temperature ``bad metal'' regime $\sigma \lesssim e^2/h$.

\section{A large $N$ limit that violates the stability bound}\label{app:largeN}

As was demonstrated recently in Ref.~\cite{boundless}, most physically useful bounds in condensed matter systems cannot be made rigorous in that it is possible to construct model problems that violate them by arbitrarily large factors.  
Indeed, one model problem discussed in that paper is a matrix large $N$ version of the Holstein model~\cite{sri,erez} for which  MEBG theory is controlled for all $\lambda \ll N$, where $N$ is the number of fermion flavors and $N^2$ is the number of independent optical phonon modes.
One consequence of this is that  polaronic instabilities are avoided, and consequently the simple relation $1/\tau_\text{tr} = 2\pi \lambda T$ remains valid, meaning that the maximum value of $\alpha$ is at least $2\pi N$, parametrically violating the proposed stability bound.

While this is an important example in principle, it is worth asking whether it is likely to be realized in physically plausible circumstances. 
The essential physics derives from the existence of a large number of phonon modes coupled in different ways (i.e.~through non-commuting fermion bilinears) to each fermionic mode.  
Consequently, it is possible to find conditions where each phonon mode is weakly coupled to the fermions, and hence not susceptible to a polaronic instability, but the net effect on the fermions can be arbitrarily strong.  

It is hard to imagine that many physically realizable systems satisfy these conditions.  
In the first place, while complex crystals do have a large number of distinct phonon modes, it is generically true that only a small number of them are coupled strongly to the Fermi-level electronic modes of the system.  
Moreover, if more than one mode is coupled to the same fermion bilinear---for instance to the local density as in the Holstein model---then one can approximately transform to a basis in which all but one combination of these modes is decoupled. 
It is plausible to have phonons coupled to more than one local fermionic density. 
For instance one could have a Holstein-like coupling to the site density and an SSH-like coupling to the bond-density~\cite{nocera2021}.  
But even here, it requires a degree of fine-tuning for the two couplings to be comparable in strength.  
Finding situations in which many such couplings are comparable would seem highly unlikely.

\section{Additional data}
\label{app:additionaldata}

In this section we present additional data demonstrating the robustness of our stated bound to the inclusion of explicit phonon anharmonicity, as well as data for other electron densities.

\begin{figure*}
	\centering
	\subfloat{%
	    \includegraphics[width=\columnwidth]{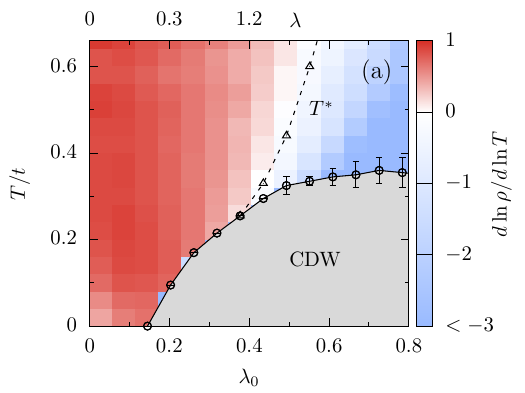}%
	}\hfill
	\subfloat{%
	    \includegraphics[width=\columnwidth]{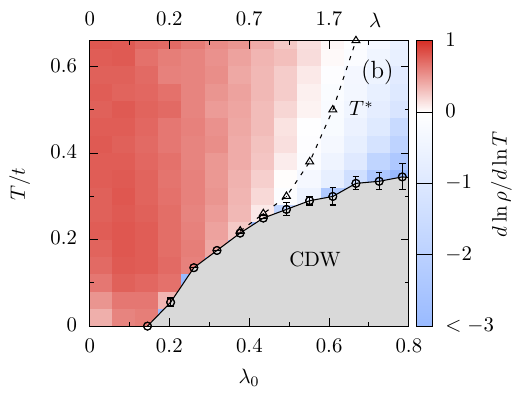}%
	}
	\caption{
	Transport phase diagram of the 3D Holstein model with static anharmonic phonons as a function of bare (renormalized) electron-phonon coupling $\lambda_0$ ($\lambda$) and temperature $T$. 
    The anharmonic potential is Eq.~\eqref{eq:quartic_potential} with $q=0.5 ~ K^2\!/t$ (panel a) and $q=2 ~ K^2\!/t$ (panel b). 
    Other parameters are the same as those used to generate the phase diagram in Fig.~1b of the main text.
	}  
	\label{fig:anharmonic_phase_diagram}
\end{figure*}

\begin{figure*}
	\centering
	\subfloat{%
	    \includegraphics[width=\columnwidth]{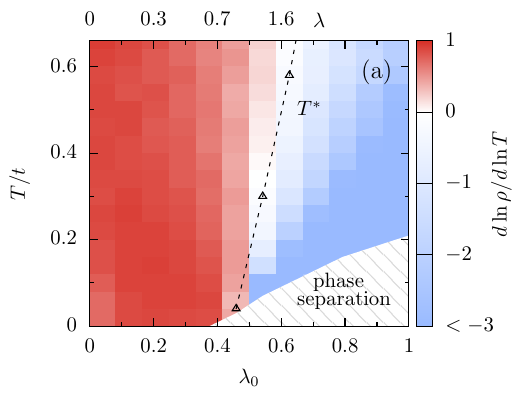}%
	}\hfill
	\subfloat{%
	    \includegraphics[width=\columnwidth]{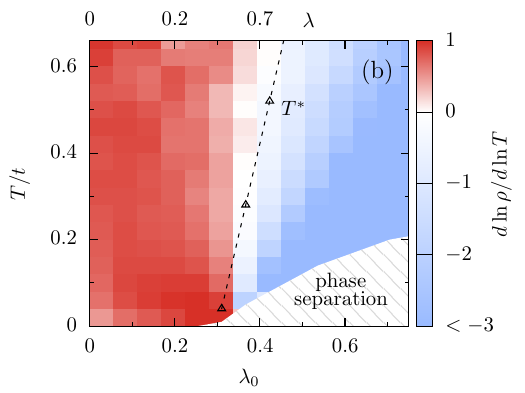}%
	}
	\caption{
	 Transport phase diagram of the 3D Holstein model with static phonons as a function of bare (renormalized) electron-phonon coupling $\lambda_0$ ($\lambda$) and temperature $T$, for electron densities $n=0.25$ (panel a) and $n=0.1$ (panel b). 
    Other parameters are the same as those used to generate the phase diagram in Fig.~1b of the main text.
	}  
	\label{fig:low-density_phase_diagram}
\end{figure*}

\subsection{Phonon anharmonicity}

A natural perturbation to add to the Holstein Hamiltonian is an anharmonic potential for the phonon; e.g.,
\begin{equation}
\label{eq:quartic_potential}
    H' = \frac{1}{4!} q x^4 \, .        
\end{equation}
The effects of such an anharmonic potential on $T_\text{CDW}$ have been studied in Ref.~\cite{Paleari:2021}. 
The relative importance of such a term will depend on the temperature and electron-phonon coupling---raising temperature increases phonon fluctuations while stronger coupling implies larger phonon distortions, both of which increase the relative importance of $H'$. 
Focusing on the latter case, we define a dimensionless measure of the strength of anharmonicity,
\begin{equation}
    r = \frac{1}{12} \frac{q U}{K^2} \, ,
\end{equation}
as the ratio of quartic to quadratic terms evaluated at a displacement $|x|=\gamma/K$, which is the displacement induced by formation of a bipolaron.
Here $U = \gamma^2/K$ is the bipolaron binding energy. 

Figure~\ref{fig:anharmonic_phase_diagram} shows the transport phase diagram in 3D with $q = 0.5 ~ K^2\!/t$ and $q = 2 ~ K^2\!/t$ (all other parameters are the same as in Fig.~1b in the main text).%
\footnote{For a harmonic phonon, coupling the phonon coordinate $x_i$ to the electron density $n_i$ is equivalent to coupling it to the density fluctuation $n_i - \bar{n}$; the difference is just a uniform shift of the phonon coordinates and chemical potential $\mu$. 
This is no longer true when anharmonic terms are included in the phonon potential. 
In general, we couple $x_i$ to $n_i - \bar{n}$.}
These values of $q$ correspond to $r = 1/24$ and $1/6$ when $U=t$ (which is near the metal-to-insulator crossover). 
There is still a critical interaction strength at which a pseudogap onsets and a sharp metal-to-insulator crossover in the resistivity. 
Importantly, the critical interaction strength, expressed in terms of the \emph{renormalized coupling}, $\lambda$, is comparable to that for vanishing bare anharmonicity, $q=0$. 
Not surprisingly, the critical \emph{bare coupling}, $\lambda_0$, differs between the two cases.

\subsection{Density dependence}

Figure~\ref{fig:low-density_phase_diagram} shows the transport phase diagram in 3D at two smaller electron densities: $n=0.25$ and $n=0.1$ (all other parameters are the same as in Fig.~1b in the main text). 
For these lower densities, there is a tendency toward phase separation at low temperatures and strong coupling. 
Despite this change in the nature of the low-temperature state of the system, we find that the metal-to-insulator crossover (and associated pseudogap onset) occurs at values of the coupling, $\lambda$, that are comparable to the values at half-filling, $n=1$, reported in the main text. 
This further demonstrates the insensitivity of the crossover to model details.

\end{document}